\pdfoutput=1
\documentclass[%
 reprint,
superscriptaddress,
nofootinbib,
 amsmath,amssymb,
 aps,
prl,
]{revtex4-1}
\usepackage{graphicx}
\usepackage{bm}
\usepackage{hyperref}
\usepackage[mathlines]{lineno}
\usepackage[utf8]{inputenc}
\usepackage{amsmath,amssymb}
\usepackage[T1]{fontenc} 
\usepackage{microtype} 
\usepackage[english]{babel} 

\usepackage{booktabs} 

\usepackage{graphicx}
\usepackage{amssymb}
\usepackage{braket,mleftright}
\usepackage{empheq}
\usepackage{subfigure}
\usepackage{stackrel}
\usepackage{soul}
\usepackage{blkarray}
\usepackage{multirow}
\usepackage{amsmath}
\usepackage{physics}
\usepackage{amsfonts}
\usepackage{bm}
\usepackage{bbold} 
\usepackage{color}
\newcommand{\beq}{\begin{equation}}
\newcommand{\eeq}{\end{equation}}
\newcommand{\bse}{\begin{subequations}}
\newcommand{\ese}{\end{subequations}}
\newcommand{\bea}{\begin{eqnarray}}
\newcommand{\eea}{\end{eqnarray}}

\usepackage[utf8]{inputenc} 
\usepackage{amsmath}

\usepackage{enumitem} 
\setlist[itemize]{noitemsep} 

\usepackage{hyperref} 
\usepackage{braket}
\usepackage{verbatim} 

\begin{document}


\title{Entangling protocols 
due to non-Markovian dynamics}

\author{Nicol\'{a}s Mirkin}
\email[mail to:]{mirkin@df.uba.ar}
\affiliation{%
Departamento de F\'{i}sica “J. J. Giambiagi” and IFIBA, FCEyN, Universidad de Buenos Aires, 1428 Buenos Aires, Argentina
}%
\author{Pablo Poggi}
\affiliation{%
Departamento de F\'{i}sica “J. J. Giambiagi” and IFIBA, FCEyN, Universidad de Buenos Aires, 1428 Buenos Aires, Argentina
}%
\affiliation{Center for Quantum Information and Control, University of New Mexico, MSC07-4220, Albuquerque, New Mexico 87131-0001, USA}
\author{Diego Wisniacki}

\affiliation{%
Departamento de F\'{i}sica “J. J. Giambiagi” and IFIBA, FCEyN, Universidad de Buenos Aires, 1428 Buenos Aires, Argentina
}%

\date{August 8, 2018}%

\begin{abstract}

It is widely spread in the literature that non-Markovianity (NM) may be regarded as a resource in quantum mechanics. However, it is still unclear how and when this alleged resource may be exploited. Here, we study the relationship between NM and quantum optimal control under the objective of generating entanglement within $M$ non-interacting subsystems, each one coupled to the same non-Markovian environment. Thus, we design a variety of entangling protocols that are only achievable due to the existence of the environment. We show that NM plays a crucial role in all the entangling protocols considered, revealing that the degree of NM completely determines the success of the entangling operation performed by the control. This is a demonstration of the virtues of NM and the way that it can be exploited in a general entangling setup.
\end{abstract}

\maketitle


\paragraph{Introduction.} The functioning of current society is based on the ability to communicate and process information. During the last two decades, it became clear that these two tasks are entering a new revolutionary era in which quantum mechanics plays a major role. One of the main challenges in this context \cite{bib:intro1,bib:intro2,bib:intro3} is the problem of controlling accurately such quantum systems. There, the study of open quantum systems is of paramount importance since any information processing device is inevitably subject to noise from the environment. Therefore, it is crucial to complete the control task in the fastest possible way in order to avoid the effects of the environment that can destroy the coherence properties of the system \cite{bib:intro_deco,bib:petru}.

However, in certain regimes known as non-Markovian, the effects of the environment may not necessarily be harmful for control, since there are memory effects that allow the flow of information from the environment back to the system, producing a momentary increase of quantum coherence. For that reason, in the last years there has been considerable interest in exploiting the phenomenon of non-Markovianity (NM) as a resource for control \cite{bib:intro_NM_res1,bib:intro_NM_res2,bib:NM_control3} and it has been argued that it can even produce an increase in the quantum speed of evolution \cite{bib:intro_speedup, bib:qsl1, bib:mirkin}. Though, besides investigations on particular systems \cite{bib:NM_control00,bib:NM_control0,bib:indios,bib:lofranco1,bib:lofranco2,bib:lofranco3,bib:pachon, bib:NM_control4,bib:haase2018controllable}, there have been no quantitative studies about controllability in non-Markovian evolutions, and currently there is very limited knowledge about which features of NM can be exploited for control as well as the connection between both \cite{bib:NM_control1,bib:NM_control2}. Several essential questions still remain unanswered. For instance, assuming the premise that NM can be considered as a resource in open quantum systems, one may be tempted to think that there should be some causal relation between the non-Markovian features of the dynamics and the controllability of the system. Furthermore, is there a control task that cannot be achieved in a Markovian regime but is attainable in a non-Markovian regime?

In this Letter we study the interplay between NM and quantum optimal control in the context of generating entanglement. We are interested in the case of an open quantum system composed by $M$ smaller subsystems that do not interact with each other but are coupled to the same non-Markovian environment. More precisely, we consider the model of a spin star configuration \cite{bib:spinstar1,bib:spinstar2,bib:spinstar3,bib:spinstar4}, where $M$ non-interacting central spins are surrounded by the same set of environmental spins. The setup was designed so that the only way of generating entanglement within the open system is due to the existence of an environment. Under this framework, we optimize a variety of entangling protocols for these non-interacting subsystems by just controlling and accessing one of them. In all the cases covered, either we perform the optimization over the space of states, i.e. starting from a separable initial multipartite state and driving the open system to a target entangled state, or either we perform the optimization over the space of gates, i.e. optimizing a field for the generation of a target entangling gate, we show that NM plays a major and essential role in the controllability of the system. In fact, we show a quantitative and direct relationship between the original degree of NM and the fidelities attained, revealing its virtues and providing a general entangling setup in which NM can be exploited.

\begin{figure}[!htb]
\begin{center}
\includegraphics[scale=0.65]{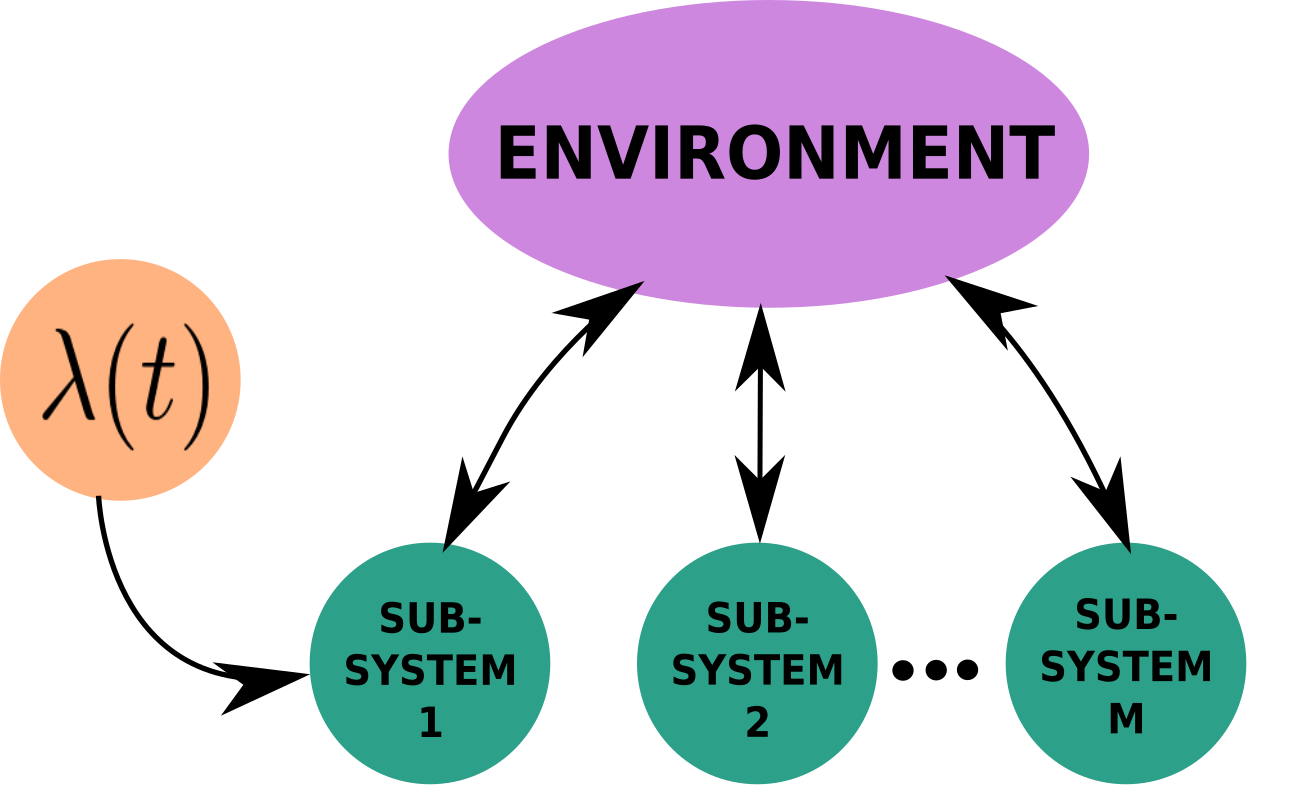}
\begin{footnotesize}
\caption{Schematic representation of the entangling setup considered under non-Markovian dynamics. The existence of a backflow of information from the environment to the system makes possible the entanglement between the subsystems by only controlling one of them. On the other hand, under a Markovian regime where information flows in one direction from the system to the environment it will not be possible to generate any arbitrary entangling transformation with accuracy.} \label{nm_ent}
\end{footnotesize}
\end{center}
\end{figure}

\paragraph{Physical model} A qualitative scheme for the entangling protocols considered is presented in Fig. \ref{nm_ent}. We stress the fact that this general physical scheme is of significant interest since it allows to study how the information from the protocol travels throughout the system within a Markovian or non-Markovian environment.
\renewcommand{\figurename}{Figure} 

As can be seen from Fig. \ref{nm_ent}, in order to generate a certain entangled state between, for example, two non-interacting subsystems, we need to give information to subsystem 2 (assuming just subsystem 1 is coupled to the control field) about the target state. This information is then communicated by the control field to subsystem 1 and from there to the environment, which only in the case of a non-Markovian regime should manage to get that information to subsystem 2. Consequently, it seems that the only way of generating controlled entanglement in this framework is due to the virtues of NM and the existence of a flow of information that goes from the environment back to the system. 

Having this qualitative argument in mind, we consider as a specific model the so-called spin star configuration. In this model we have $M$ central spin-$\frac{1}{2}$ particles that are surrounded by a set of $N-M$ likewise particles \cite{bib:spinstar1,bib:spinstar2,bib:spinstar3,bib:spinstar4}. From now on, we will refer to the surrounding spins as the environmental spins and to the central spins as the open system. We assume that the central spins do not interact with each other, that they are subjected to a constant magnetic field in the $\hat{z}$ direction, and that they additionally interact with the environmental spins via an isotropic Heisenberg interaction. We stress the fact that just one of the central spins can be accessed and controlled via a time-dependent control field in the $\hat{y}$ direction. The dynamics of this model is then described by the following Hamiltonian   
\begin{equation}
\begin{split}
    H & = H_{0}+H_{C}(t) \\
    & = \sum_{l=0}^{M-1} \dfrac{\omega_{0}}{2}\sigma_{z}^{(l)}+\sum_{l=0}^{M-1}\sum_{k=M}^{N-1} \left( A_{k} \vec{\sigma}^{(l)}.\vec{\sigma}^{(k)} \right) +\lambda(t)\sigma_{y}^{(0)},
\end{split}
\end{equation}
where $H_{0}$ plays the role of the free Hamiltonian and $H_{C}(t)$ is the control Hamiltonian (note that we set $\hbar=1$ from now on). As can be seen, the central spins have an energy splitting $\omega_{0}$ and are coupled to the $k$-th environmental spin via the coupling constant $A_{k}$. The operators $\vec{\sigma}^{(l)}$ and $\vec{\sigma}^{(k)}$ are the Pauli operators of the $l$-th central spin and the $k$-th environmental spin, respectively, and the quantity $\lambda(t)$ is the control field with which we will drive our open system to the desired target. Works that have sought to control a similar system can be found in Refs. \cite{bib:controlarenz, bib:controlfloether}, but none has directly related NM, entanglement and optimal control as we propose here. 
Imposing equal system-environment couplings ($A_{k}=A$ for each k), considering one central spin (M=1), and neglecting the control Hamiltonian $H_{C}(t)$, one can derive the exact reduced dynamics for the spin star, as has been previously shown in Refs. \cite{bib:spinstar1,bib:spinstar2,bib:spinstar3,bib:spinstar4}. A significant fact is that the interaction between a central spin and a bath of environmental spins leads to an intrinsically non-Markovian behavior. However, in order to describe the full reduced dynamics of the controlled system, one would have to derive a non-Markovian master equation that depends explicitly on the unknown field $\lambda(t)$ and the exact solution to this remains an open and a challenging problem in the context of controlling general open quantum systems \cite{bib:controlopen0, bib:controlopen1,bib:controlopen2}. For the purposes of this Letter, we will restrict ourselves to the case of an environment formed by a few spins and where we have complete knowledge over the total dynamics of the system. In this sense, even though we are just interested in the open dynamics of the central spins, our framework will be the Schrödinger equation and after solving the whole unitary optimized evolution we will have to trace over the environmental spins to obtain the sought reduced dynamics. In this context we will address the interplay between quantum optimal control and the NM of the system dynamics. 

\paragraph{Entangling protocols due to non-Markovian dynamics.} As we are interested in driving the $M$ central spins of the open system from an initial separable state to a target entangled state, we need to resort to numerical state optimization where we optimize over an initial random field in order to maximize the state fidelity defined as $\mathcal{F}_{state}=|\bra{\psi_{targ}}\ket{\psi(T)}|^{2}$ \cite{bib:qutip}. The driving time $T$ has been divided into 200 equidistant time intervals, enough to ensure a proper resolution of the dynamics. But to proceed we still need to define how we will quantify the non-Markovian features of the dynamics. 

During the last decade there have been several proposals to describe and quantify non-Markovian effects in open quantum systems \cite{bib:nmreport,bib:manis,bib:pineda,bib:pablo}. One of the most popular approaches was given by Breuer, Laine and Piilo (BLP) \cite{bib:nmbreuer1,bib:nmbreuer2,bib:wissmann}, who based their measure in the revivals of distinguishability between quantum states during the dynamics. For this approach is necessary to define a measure of distance and distinguishability between two states. The so-called trace distance is defined as $D(\rho_{1},\rho_{2})=\dfrac{1}{2}||\rho_{1}-\rho_{2}||$,
where $||A||=tr(\sqrt{A^{\dagger}A})$. Thus, the BLP criterion states that a quantum map is non-Markovian if there exists at least a pair of initial states $\rho_{1}(0)$ and $\rho_{2}(0)$ such that $\sigma(\rho_{1}(0),\rho_{2}(0),t)=\dfrac{d}{dt}D(\rho_{1}(t),\rho_{2}(t)) > 0$,
for some interval of time. The physical meaning of the above is that the states $\rho_{1}(t)$ and $\rho_{2}(t)$ are becoming momentarily more distinguishable and this is equivalent to say that information has flowed from the environment back to the system. Instead, in Markovian dynamics information is continuously lost to the environment and the map ultimately bears no memory of the initial state of the system. In this way, by quantifying the total amount of information backflow during the evolution, the BLP criterion can also be extended to define a measure of the degree of NM in a quantum process via 
\begin{equation}
    \mathcal{N}=\max\limits_{{\lbrace\rho_{1}(0),\rho_{2}(0)\rbrace}} \int_{0, \sigma >0}^{T}\sigma \left (\rho_{1}(0),\rho_{2}(0),t'\right) dt',
    \label{BLP}
\end{equation}
where $T$ stands for the final evolution time of the process under consideration. We stress the fact that the NM in this Letter is quantified for a restricted time interval, due to considering a finite evolution time for the control protocol, which may be varied.
As the initial states for Eq. (\ref{BLP}) we take the separable states
\begin{equation}
\ket{\psi_{1,2}(0)}_{S\otimes E} =\bigotimes\limits_{l=1}^{M}\dfrac{1}{\sqrt{2}}(\ket{0}\pm\ket{1})_{S_{l}} \bigotimes\limits_{n=1}^{N-M}\ket{1}_{E_{n}},
\label{in_states}
\end{equation}
where $S_{l}$ stands for the state of the l-th central spin and $E_{n}$ for the n-th environmental spin, respectively. The initial states of the open system have been chosen as to be orthogonal. The procedure consists on solving the Schrödinger equation for both states of Eq. (\ref{in_states}) and after having the whole evolution then trace over the spins of the environment in order to obtain the reduced open dynamics of the central spins. With this recipe we are able to measure the distinguishability of the initial states along the whole evolution and in consequence witnessing the NM of the free process. 

\begin{figure}[!htb]
\begin{center}
\includegraphics[scale=0.7]{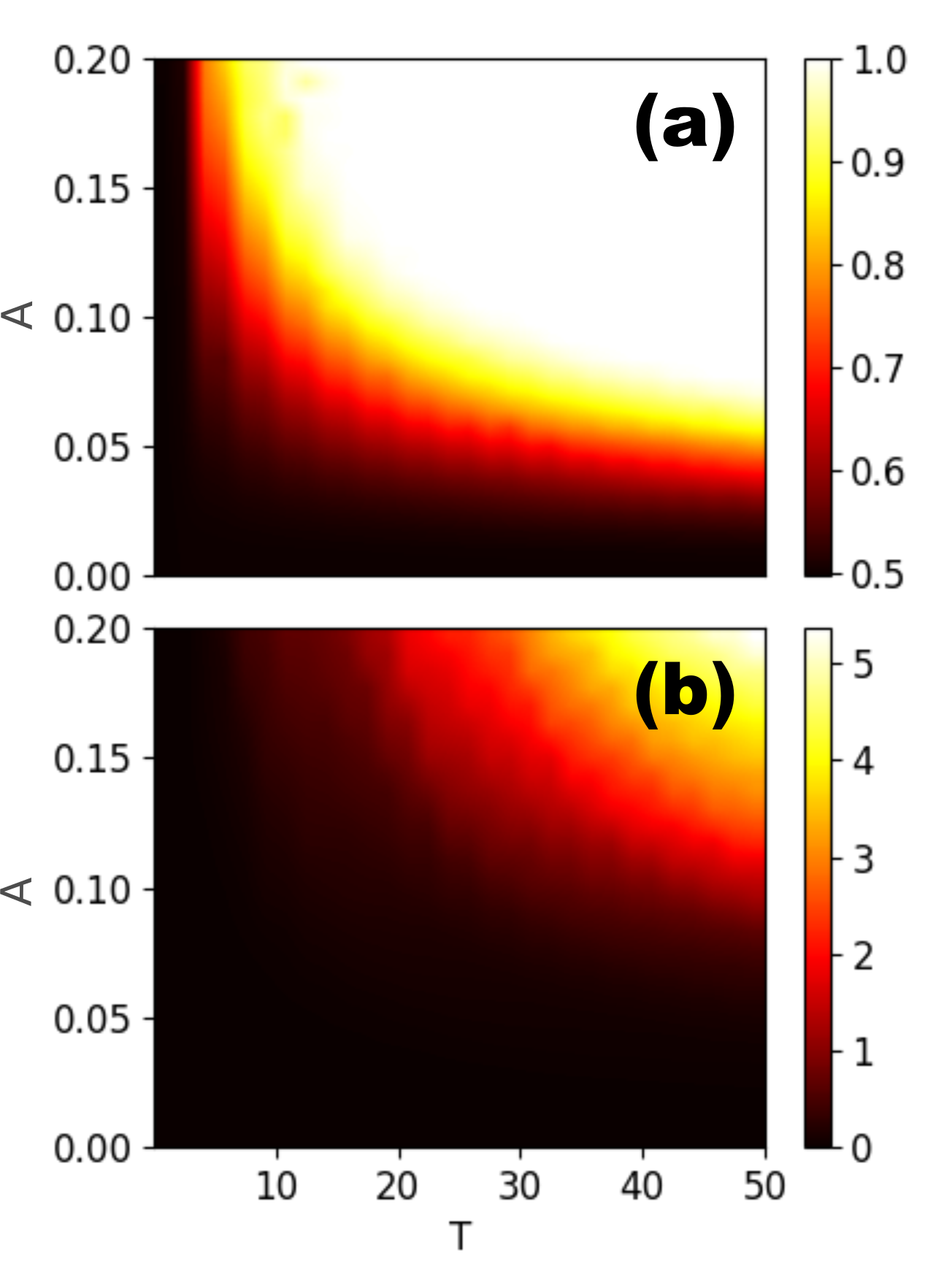}
\begin{footnotesize}
\caption{Driving to a GHZ state (M=3). (a) Fidelity achieved in the space of parameters given by the coupling A and the driving time T ($\hbar=1$ and in units of $\omega_{0}$). (b) NM of the free evolution.} 
\label{fig2}
\end{footnotesize}
\end{center}
\end{figure}

Let us start considering a configuration in which we have three spins acting as the open system ($M=3$) with two extra spins acting as the environment ($N-M=2$), where we intend to drive the three central spins from the separable state in Eq. (\ref{in_states}) to an entangled GHZ state. In Fig. \ref{fig2} we show the final fidelities achieved for this entangling protocol as well as the original degree of NM, both as a function of coupling and the duration of the process. Therefore, following one of the motor questions of this Letter, we can see that the region where the free system was originally more non-Markovian coincides with the region where the system proved to be more controllable, at least for our specific task of generating entanglement. We remark that this result was obtained for all other optimization in which we pursue the generation of entanglement between the central spins, either driving the open system to an entangled target state of the form of a Bell (M=2) or W state (M=3), or even running the optimization for implementing a target entangling gate with M=2 or M=3 (not shown). Nevertheless, this is merely a correlation between the control fidelities and the original amount of NM, both as a function of coupling and time.
\renewcommand{\figurename}{Figure} 

\begin{figure}[!htb]
\begin{center}
\includegraphics[scale=0.67]{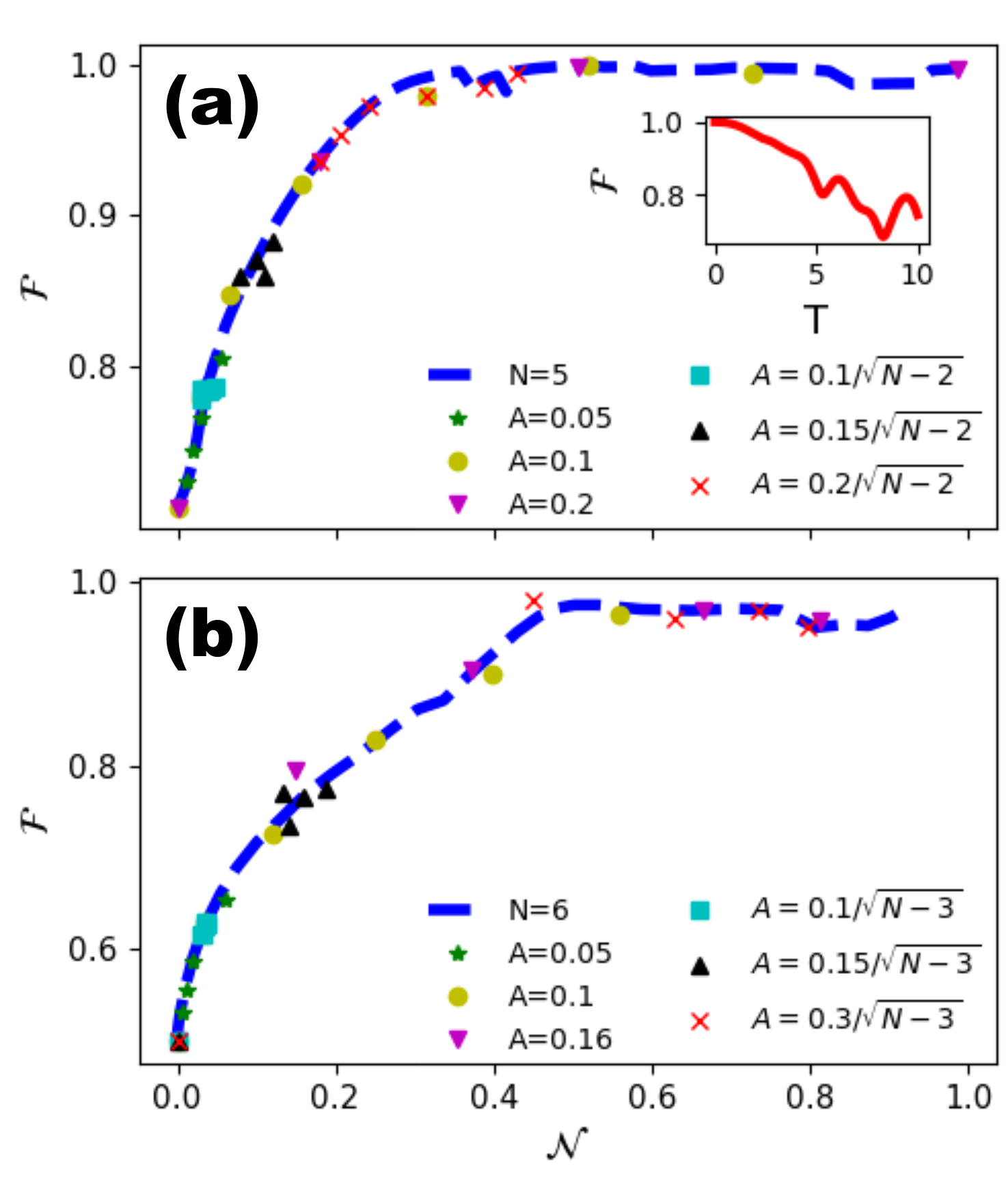}
\begin{footnotesize}
\caption{(a) M=2: driving to a Bell state. (b) M=3: driving to a GHZ state. In both panels we plot the fidelity (with control) as a function of the original NM, merging seven different protocols which have arbitrarily different couplings and number of environmental spins. While the blue dashed curve corresponds to a dynamics in which N is fixed but the coupling varies ($0\leq A \leq 0.2$), in the six other cases coupling is fixed but N varies ($M\leq N \leq 8$). In all protocols, evolution time is fixed in T=10. The inset of the upper panel shows the fidelity (without control) as a function of time between the same initial state of the open system (M=2) evolved with two different dynamics with the same original degree of non-Markovianity (NM=0.43). There we consider N=8 and $A=0.2/\sqrt{6}$ (i.e. the last red cross from panel (a)) while on the other hand we take N=5 and A=0.1466.}   
\label{bell_nm}
\end{footnotesize}
\end{center}
\end{figure}

So another straightforward thing one could do is to fix coupling strength and driving time, and analyze whether there also exists a correlation between both quantities as a function of the number of environmental spins. However, in this case one could argue that increasing the size of the environment at a fixed coupling is somehow like strengthening the interaction and so equivalent to increasing the coupling. For that reason, a possibility is to work with a fixed unscaled coupling $A$ but also with a scaled coupling parameter of the form $A'=A/\sqrt{N-M}$ in order to fix the strength of the interaction \cite{bib:spinstar2}. To obtain a complete assessment of how all the variables in the model impact the interplay between fidelity and NM, we study several configurations considering arbitrarily different couplings and number of environmental spins. Such analysis is presented in Fig. \ref{bell_nm}, where we merely focus on the fidelities attained as a function of the amount of NM of the free process. We study two different configurations, one with two central spins (M=2) where the goal is to drive the open system from a separable initial state to an entangled Bell state, and another with three central spins (M=3) where the target is an entangled GHZ state. Note that in both situations we consider several different dynamics, either varying the number of environmental spins with different but fixed couplings or either varying the coupling with a fixed N.
\renewcommand{\figurename}{Figure} 

Remarkably, we can see from both panels in Fig. \ref{bell_nm} that whatever is the system dynamics we consider, that is to say, whatever the coupling and the number of environmental spins, in each configuration the same original amount of NM leads to the same entangling fidelities. Therefore, more than correlations, what we are showing is that the entangling fidelity is a direct function of NM, revealing a causal relationship between the controllability of the system and the NM of the dynamics. This is somewhat surprising since one might well say that the reason why with two different couplings and number of environmental spins we have the same amount of NM is because the interaction between the open system and the environment is also the same. But in the inset of the upper panel in Fig. \ref{bell_nm} we have clearly shown that this is not the case. There we plot the fidelity (without control) as a function of time between the same initial state of the open system evolved by two different dynamics ($N=8$, $A=0.2/\sqrt{6}$ and $N=5$, $A=0.1466$), both with same original amount of NM, resulting that the free evolution of the open system is quite different in the two cases, despite rendering similar fidelities in the controlled case. In this way, we are demonstrating in a practical situation that there exists a causal and strong relation between the original amount of NM and the fidelities attained, independently of the parameters of the problem. Consequently, as can be deduced from Fig. \ref{bell_nm}, when the system dynamics originally had some little amount of NM, this little amount leads to a significant increment of the fidelity. In the same way, as NM continues to increase fidelity does too, although with a lower slope than it did initially. This behaviour seems reasonable since, as already mentioned, the central spins do not interact with each other and we can only control and access one of them. For this reason, is not only necessary to have an environment in order to fulfill these variety of entangling protocols but is also a necessary condition the existence of a flow of information from the environment back to the system, as can be qualitatively seen from the scheme in Fig. \ref{nm_ent} and quantitatively supported by all our calculations. 

\paragraph{Concluding remarks.} In this Letter we have aimed to deepen our understanding of the interplay between NM and quantum optimal control in the context of generating entanglement. For this purpose, we have considered a general physical scheme of an open quantum system composed by $M$ non-interacting subsystems, each one coupled to the same non-Markovian environment. In this framework, we have presented not only a qualitative interpretation about the physical mechanism underlying the generation of controlled entanglement, but also quantitative evidence revealing a causal relationship between the entangling fidelities and the original degree of NM of the system dynamics. In this sense, this Letter consists in a practical demonstration of the virtues of NM, providing a general physical setup in which it can be exploited. 

With the results in our backs, we sincerely hope this work will serve as a springboard towards a greater degree of understanding between NM, entanglement and quantum optimal control.

\begin{acknowledgements}
N.M. acknowledges Ian Petersen for fruitful discussions during PRACQSYS 2018. This work was partially supported by CONICET (PIP 112201 501004 93CO), UBACyT (20020130100406BA), ANPCyT (PICT-2016-1056)  and National Science Foundation (Grant No. PHY-1630114).
\end{acknowledgements}

\bibliography{main.bib}

\begin{thebibliography}{39}
\expandafter\ifx\csname natexlab\endcsname\relax\def\natexlab#1{#1}\fi
\expandafter\ifx\csname bibnamefont\endcsname\relax
  \def\bibnamefont#1{#1}\fi
\expandafter\ifx\csname bibfnamefont\endcsname\relax
  \def\bibfnamefont#1{#1}\fi
\expandafter\ifx\csname citenamefont\endcsname\relax
  \def\citenamefont#1{#1}\fi
\expandafter\ifx\csname url\endcsname\relax
  \def\url#1{\texttt{#1}}\fi
\expandafter\ifx\csname urlprefix\endcsname\relax\def\urlprefix{URL }\fi
\providecommand{\bibinfo}[2]{#2}
\providecommand{\eprint}[2][]{\url{#2}}

\bibitem[{\citenamefont{Bekenstein}(1981)}]{bib:intro1}
\bibinfo{author}{\bibfnamefont{J.~D.} \bibnamefont{Bekenstein}},
  \bibinfo{journal}{Phys. Rev. Lett.} \textbf{\bibinfo{volume}{46}},
  \bibinfo{pages}{623} (\bibinfo{year}{1981}).

\bibitem[{\citenamefont{Lloyd}(2000)}]{bib:intro2}
\bibinfo{author}{\bibfnamefont{S.}~\bibnamefont{Lloyd}},
  \bibinfo{journal}{Nature} \textbf{\bibinfo{volume}{406}},
  \bibinfo{pages}{1047} (\bibinfo{year}{2000}).

\bibitem[{\citenamefont{Giovannetti et~al.}(2011)\citenamefont{Giovannetti,
  Lloyd, and Maccone}}]{bib:intro3}
\bibinfo{author}{\bibfnamefont{V.}~\bibnamefont{Giovannetti}},
  \bibinfo{author}{\bibfnamefont{S.}~\bibnamefont{Lloyd}}, \bibnamefont{and}
  \bibinfo{author}{\bibfnamefont{L.}~\bibnamefont{Maccone}},
  \bibinfo{journal}{Nature photonics} \textbf{\bibinfo{volume}{5}},
  \bibinfo{pages}{222} (\bibinfo{year}{2011}).

\bibitem[{\citenamefont{Schlosshauer}(2007)}]{bib:intro_deco}
\bibinfo{author}{\bibfnamefont{M.~A.} \bibnamefont{Schlosshauer}},
  \emph{\bibinfo{title}{Decoherence: and the quantum-to-classical transition}}
  (\bibinfo{publisher}{Springer Science \& Business Media},
  \bibinfo{year}{2007}).

\bibitem[{\citenamefont{Breuer et~al.}(2002)\citenamefont{Breuer, Petruccione
  et~al.}}]{bib:petru}
\bibinfo{author}{\bibfnamefont{H.-P.} \bibnamefont{Breuer}},
  \bibinfo{author}{\bibfnamefont{F.}~\bibnamefont{Petruccione}},
  \bibnamefont{et~al.}, \emph{\bibinfo{title}{The theory of open quantum
  systems}} (\bibinfo{publisher}{Oxford University Press on Demand},
  \bibinfo{year}{2002}).

\bibitem[{\citenamefont{Breuer}(2012)}]{bib:intro_NM_res1}
\bibinfo{author}{\bibfnamefont{H.-P.} \bibnamefont{Breuer}},
  \bibinfo{journal}{Journal of Physics B: Atomic, Molecular and Optical
  Physics} \textbf{\bibinfo{volume}{45}}, \bibinfo{pages}{154001}
  (\bibinfo{year}{2012}).

\bibitem[{\citenamefont{Bylicka et~al.}(2014)\citenamefont{Bylicka,
  Chru{\'s}ci{\'n}ski, and Maniscalco}}]{bib:intro_NM_res2}
\bibinfo{author}{\bibfnamefont{B.}~\bibnamefont{Bylicka}},
  \bibinfo{author}{\bibfnamefont{D.}~\bibnamefont{Chru{\'s}ci{\'n}ski}},
  \bibnamefont{and}
  \bibinfo{author}{\bibfnamefont{S.}~\bibnamefont{Maniscalco}},
  \bibinfo{journal}{Scientific reports} \textbf{\bibinfo{volume}{4}}
  (\bibinfo{year}{2014}).

\bibitem[{\citenamefont{Reich et~al.}(2015)\citenamefont{Reich, Katz, and
  Koch}}]{bib:NM_control3}
\bibinfo{author}{\bibfnamefont{D.~M.} \bibnamefont{Reich}},
  \bibinfo{author}{\bibfnamefont{N.}~\bibnamefont{Katz}}, \bibnamefont{and}
  \bibinfo{author}{\bibfnamefont{C.~P.} \bibnamefont{Koch}},
  \bibinfo{journal}{Scientific reports} \textbf{\bibinfo{volume}{5}}
  (\bibinfo{year}{2015}).

\bibitem[{\citenamefont{Cimmarusti et~al.}(2015)\citenamefont{Cimmarusti, Yan,
  Patterson, Corcos, Orozco, and Deffner}}]{bib:intro_speedup}
\bibinfo{author}{\bibfnamefont{A.}~\bibnamefont{Cimmarusti}},
  \bibinfo{author}{\bibfnamefont{Z.}~\bibnamefont{Yan}},
  \bibinfo{author}{\bibfnamefont{B.}~\bibnamefont{Patterson}},
  \bibinfo{author}{\bibfnamefont{L.}~\bibnamefont{Corcos}},
  \bibinfo{author}{\bibfnamefont{L.}~\bibnamefont{Orozco}}, \bibnamefont{and}
  \bibinfo{author}{\bibfnamefont{S.}~\bibnamefont{Deffner}},
  \bibinfo{journal}{Physical review letters} \textbf{\bibinfo{volume}{114}},
  \bibinfo{pages}{233602} (\bibinfo{year}{2015}).

\bibitem[{\citenamefont{Deffner and Lutz}(2013)}]{bib:qsl1}
\bibinfo{author}{\bibfnamefont{S.}~\bibnamefont{Deffner}} \bibnamefont{and}
  \bibinfo{author}{\bibfnamefont{E.}~\bibnamefont{Lutz}},
  \bibinfo{journal}{Physical review letters} \textbf{\bibinfo{volume}{111}},
  \bibinfo{pages}{010402} (\bibinfo{year}{2013}).

\bibitem[{\citenamefont{Mirkin et~al.}(2016)\citenamefont{Mirkin, Toscano, and
  Wisniacki}}]{bib:mirkin}
\bibinfo{author}{\bibfnamefont{N.}~\bibnamefont{Mirkin}},
  \bibinfo{author}{\bibfnamefont{F.}~\bibnamefont{Toscano}}, \bibnamefont{and}
  \bibinfo{author}{\bibfnamefont{D.~A.} \bibnamefont{Wisniacki}},
  \bibinfo{journal}{Physical Review A} \textbf{\bibinfo{volume}{94}},
  \bibinfo{pages}{052125} (\bibinfo{year}{2016}).

\bibitem[{\citenamefont{Schmidt
  et~al.}(2011{\natexlab{a}})\citenamefont{Schmidt, Negretti, Ankerhold,
  Calarco, and Stockburger}}]{bib:NM_control00}
\bibinfo{author}{\bibfnamefont{R.}~\bibnamefont{Schmidt}},
  \bibinfo{author}{\bibfnamefont{A.}~\bibnamefont{Negretti}},
  \bibinfo{author}{\bibfnamefont{J.}~\bibnamefont{Ankerhold}},
  \bibinfo{author}{\bibfnamefont{T.}~\bibnamefont{Calarco}}, \bibnamefont{and}
  \bibinfo{author}{\bibfnamefont{J.~T.} \bibnamefont{Stockburger}},
  \bibinfo{journal}{Physical review letters} \textbf{\bibinfo{volume}{107}},
  \bibinfo{pages}{130404} (\bibinfo{year}{2011}{\natexlab{a}}).

\bibitem[{\citenamefont{Hwang and Goan}(2012)}]{bib:NM_control0}
\bibinfo{author}{\bibfnamefont{B.}~\bibnamefont{Hwang}} \bibnamefont{and}
  \bibinfo{author}{\bibfnamefont{H.-S.} \bibnamefont{Goan}},
  \bibinfo{journal}{Physical Review A} \textbf{\bibinfo{volume}{85}},
  \bibinfo{pages}{032321} (\bibinfo{year}{2012}).

\bibitem[{\citenamefont{Khurana et~al.}(2018)\citenamefont{Khurana, Agarwalla,
  and Mahesh}}]{bib:indios}
\bibinfo{author}{\bibfnamefont{D.}~\bibnamefont{Khurana}},
  \bibinfo{author}{\bibfnamefont{B.~K.} \bibnamefont{Agarwalla}},
  \bibnamefont{and} \bibinfo{author}{\bibfnamefont{T.}~\bibnamefont{Mahesh}},
  \bibinfo{journal}{arXiv preprint arXiv:1805.10772}  (\bibinfo{year}{2018}).

\bibitem[{\citenamefont{Franco et~al.}(2013)\citenamefont{Franco, Bellomo,
  Maniscalco, and Compagno}}]{bib:lofranco1}
\bibinfo{author}{\bibfnamefont{R.~L.} \bibnamefont{Franco}},
  \bibinfo{author}{\bibfnamefont{B.}~\bibnamefont{Bellomo}},
  \bibinfo{author}{\bibfnamefont{S.}~\bibnamefont{Maniscalco}},
  \bibnamefont{and} \bibinfo{author}{\bibfnamefont{G.}~\bibnamefont{Compagno}},
  \bibinfo{journal}{International Journal of Modern Physics B}
  \textbf{\bibinfo{volume}{27}}, \bibinfo{pages}{1345053}
  (\bibinfo{year}{2013}).

\bibitem[{\citenamefont{Man et~al.}(2015)\citenamefont{Man, Xia, and
  Franco}}]{bib:lofranco2}
\bibinfo{author}{\bibfnamefont{Z.-X.} \bibnamefont{Man}},
  \bibinfo{author}{\bibfnamefont{Y.-J.} \bibnamefont{Xia}}, \bibnamefont{and}
  \bibinfo{author}{\bibfnamefont{R.~L.} \bibnamefont{Franco}},
  \bibinfo{journal}{Physical Review A} \textbf{\bibinfo{volume}{92}},
  \bibinfo{pages}{012315} (\bibinfo{year}{2015}).

\bibitem[{\citenamefont{Bellomo et~al.}(2007)\citenamefont{Bellomo, Franco, and
  Compagno}}]{bib:lofranco3}
\bibinfo{author}{\bibfnamefont{B.}~\bibnamefont{Bellomo}},
  \bibinfo{author}{\bibfnamefont{R.~L.} \bibnamefont{Franco}},
  \bibnamefont{and} \bibinfo{author}{\bibfnamefont{G.}~\bibnamefont{Compagno}},
  \bibinfo{journal}{Physical Review Letters} \textbf{\bibinfo{volume}{99}},
  \bibinfo{pages}{160502} (\bibinfo{year}{2007}).

\bibitem[{\citenamefont{Estrada and Pach{\'o}n}(2015)}]{bib:pachon}
\bibinfo{author}{\bibfnamefont{A.~F.} \bibnamefont{Estrada}} \bibnamefont{and}
  \bibinfo{author}{\bibfnamefont{L.~A.} \bibnamefont{Pach{\'o}n}},
  \bibinfo{journal}{New Journal of Physics} \textbf{\bibinfo{volume}{17}},
  \bibinfo{pages}{033038} (\bibinfo{year}{2015}).

\bibitem[{\citenamefont{Torre and Illuminati}(2018)}]{bib:NM_control4}
\bibinfo{author}{\bibfnamefont{G.}~\bibnamefont{Torre}} \bibnamefont{and}
  \bibinfo{author}{\bibfnamefont{F.}~\bibnamefont{Illuminati}},
  \bibinfo{journal}{arXiv preprint arXiv:1805.03617}  (\bibinfo{year}{2018}).

\bibitem[{\citenamefont{Haase et~al.}(2018)\citenamefont{Haase, Vetter, Unden,
  Smirne, Rosskopf, Naydenov, Jelezko, Plenio, and
  Huelga}}]{bib:haase2018controllable}
\bibinfo{author}{\bibfnamefont{J.~F.} \bibnamefont{Haase}},
  \bibinfo{author}{\bibfnamefont{P.~J.} \bibnamefont{Vetter}},
  \bibinfo{author}{\bibfnamefont{T.}~\bibnamefont{Unden}},
  \bibinfo{author}{\bibfnamefont{A.}~\bibnamefont{Smirne}},
  \bibinfo{author}{\bibfnamefont{J.}~\bibnamefont{Rosskopf}},
  \bibinfo{author}{\bibfnamefont{B.}~\bibnamefont{Naydenov}},
  \bibinfo{author}{\bibfnamefont{F.}~\bibnamefont{Jelezko}},
  \bibinfo{author}{\bibfnamefont{M.~B.} \bibnamefont{Plenio}},
  \bibnamefont{and} \bibinfo{author}{\bibfnamefont{S.~F.}
  \bibnamefont{Huelga}}, \bibinfo{journal}{arXiv preprint arXiv:1802.00819}
  (\bibinfo{year}{2018}).

\bibitem[{\citenamefont{Glaser et~al.}(2015)\citenamefont{Glaser, Boscain,
  Calarco, Koch, K{\"o}ckenberger, Kosloff, Kuprov, Luy, Schirmer,
  Schulte-Herbr{\"u}ggen et~al.}}]{bib:NM_control1}
\bibinfo{author}{\bibfnamefont{S.~J.} \bibnamefont{Glaser}},
  \bibinfo{author}{\bibfnamefont{U.}~\bibnamefont{Boscain}},
  \bibinfo{author}{\bibfnamefont{T.}~\bibnamefont{Calarco}},
  \bibinfo{author}{\bibfnamefont{C.~P.} \bibnamefont{Koch}},
  \bibinfo{author}{\bibfnamefont{W.}~\bibnamefont{K{\"o}ckenberger}},
  \bibinfo{author}{\bibfnamefont{R.}~\bibnamefont{Kosloff}},
  \bibinfo{author}{\bibfnamefont{I.}~\bibnamefont{Kuprov}},
  \bibinfo{author}{\bibfnamefont{B.}~\bibnamefont{Luy}},
  \bibinfo{author}{\bibfnamefont{S.}~\bibnamefont{Schirmer}},
  \bibinfo{author}{\bibfnamefont{T.}~\bibnamefont{Schulte-Herbr{\"u}ggen}},
  \bibnamefont{et~al.}, \bibinfo{journal}{The European Physical Journal D}
  \textbf{\bibinfo{volume}{69}}, \bibinfo{pages}{279} (\bibinfo{year}{2015}).

\bibitem[{\citenamefont{Koch}(2016)}]{bib:NM_control2}
\bibinfo{author}{\bibfnamefont{C.~P.} \bibnamefont{Koch}},
  \bibinfo{journal}{Journal of Physics: Condensed Matter}
  \textbf{\bibinfo{volume}{28}}, \bibinfo{pages}{213001}
  (\bibinfo{year}{2016}).

\bibitem[{\citenamefont{Hutton and Bose}(2004)}]{bib:spinstar1}
\bibinfo{author}{\bibfnamefont{A.}~\bibnamefont{Hutton}} \bibnamefont{and}
  \bibinfo{author}{\bibfnamefont{S.}~\bibnamefont{Bose}},
  \bibinfo{journal}{Physical Review A} \textbf{\bibinfo{volume}{69}},
  \bibinfo{pages}{042312} (\bibinfo{year}{2004}).

\bibitem[{\citenamefont{Breuer et~al.}(2004)\citenamefont{Breuer, Burgarth, and
  Petruccione}}]{bib:spinstar2}
\bibinfo{author}{\bibfnamefont{H.-P.} \bibnamefont{Breuer}},
  \bibinfo{author}{\bibfnamefont{D.}~\bibnamefont{Burgarth}}, \bibnamefont{and}
  \bibinfo{author}{\bibfnamefont{F.}~\bibnamefont{Petruccione}},
  \bibinfo{journal}{Physical Review B} \textbf{\bibinfo{volume}{70}},
  \bibinfo{pages}{045323} (\bibinfo{year}{2004}).

\bibitem[{\citenamefont{Fischer and Breuer}(2007)}]{bib:spinstar3}
\bibinfo{author}{\bibfnamefont{J.}~\bibnamefont{Fischer}} \bibnamefont{and}
  \bibinfo{author}{\bibfnamefont{H.-P.} \bibnamefont{Breuer}},
  \bibinfo{journal}{Physical Review A} \textbf{\bibinfo{volume}{76}},
  \bibinfo{pages}{052119} (\bibinfo{year}{2007}).

\bibitem[{\citenamefont{Semin et~al.}(2014)\citenamefont{Semin, Sinayskiy, and
  Petruccione}}]{bib:spinstar4}
\bibinfo{author}{\bibfnamefont{V.}~\bibnamefont{Semin}},
  \bibinfo{author}{\bibfnamefont{I.}~\bibnamefont{Sinayskiy}},
  \bibnamefont{and}
  \bibinfo{author}{\bibfnamefont{F.}~\bibnamefont{Petruccione}},
  \bibinfo{journal}{Physical Review A} \textbf{\bibinfo{volume}{89}},
  \bibinfo{pages}{012107} (\bibinfo{year}{2014}).

\bibitem[{\citenamefont{Arenz et~al.}(2014)\citenamefont{Arenz, Gualdi, and
  Burgarth}}]{bib:controlarenz}
\bibinfo{author}{\bibfnamefont{C.}~\bibnamefont{Arenz}},
  \bibinfo{author}{\bibfnamefont{G.}~\bibnamefont{Gualdi}}, \bibnamefont{and}
  \bibinfo{author}{\bibfnamefont{D.}~\bibnamefont{Burgarth}},
  \bibinfo{journal}{New Journal of Physics} \textbf{\bibinfo{volume}{16}},
  \bibinfo{pages}{065023} (\bibinfo{year}{2014}).

\bibitem[{\citenamefont{Floether et~al.}(2012)\citenamefont{Floether,
  de~Fouquieres, and Schirmer}}]{bib:controlfloether}
\bibinfo{author}{\bibfnamefont{F.~F.} \bibnamefont{Floether}},
  \bibinfo{author}{\bibfnamefont{P.}~\bibnamefont{de~Fouquieres}},
  \bibnamefont{and} \bibinfo{author}{\bibfnamefont{S.~G.}
  \bibnamefont{Schirmer}}, \bibinfo{journal}{New Journal of Physics}
  \textbf{\bibinfo{volume}{14}}, \bibinfo{pages}{073023}
  (\bibinfo{year}{2012}).

\bibitem[{\citenamefont{Schmidt
  et~al.}(2011{\natexlab{b}})\citenamefont{Schmidt, Negretti, Ankerhold,
  Calarco, and Stockburger}}]{bib:controlopen0}
\bibinfo{author}{\bibfnamefont{R.}~\bibnamefont{Schmidt}},
  \bibinfo{author}{\bibfnamefont{A.}~\bibnamefont{Negretti}},
  \bibinfo{author}{\bibfnamefont{J.}~\bibnamefont{Ankerhold}},
  \bibinfo{author}{\bibfnamefont{T.}~\bibnamefont{Calarco}}, \bibnamefont{and}
  \bibinfo{author}{\bibfnamefont{J.~T.} \bibnamefont{Stockburger}},
  \bibinfo{journal}{Physical review letters} \textbf{\bibinfo{volume}{107}},
  \bibinfo{pages}{130404} (\bibinfo{year}{2011}{\natexlab{b}}).

\bibitem[{\citenamefont{O'Meara et~al.}(2012)\citenamefont{O'Meara, Dirr, and
  Schulte-Herbruggen}}]{bib:controlopen1}
\bibinfo{author}{\bibfnamefont{C.}~\bibnamefont{O'Meara}},
  \bibinfo{author}{\bibfnamefont{G.}~\bibnamefont{Dirr}}, \bibnamefont{and}
  \bibinfo{author}{\bibfnamefont{T.}~\bibnamefont{Schulte-Herbruggen}},
  \bibinfo{journal}{IEEE Transactions on Automatic Control}
  \textbf{\bibinfo{volume}{57}}, \bibinfo{pages}{2050} (\bibinfo{year}{2012}).

\bibitem[{\citenamefont{Schulte-Herbr{\"u}ggen
  et~al.}(2011)\citenamefont{Schulte-Herbr{\"u}ggen, Sp{\"o}rl, Khaneja, and
  Glaser}}]{bib:controlopen2}
\bibinfo{author}{\bibfnamefont{T.}~\bibnamefont{Schulte-Herbr{\"u}ggen}},
  \bibinfo{author}{\bibfnamefont{A.}~\bibnamefont{Sp{\"o}rl}},
  \bibinfo{author}{\bibfnamefont{N.}~\bibnamefont{Khaneja}}, \bibnamefont{and}
  \bibinfo{author}{\bibfnamefont{S.}~\bibnamefont{Glaser}},
  \bibinfo{journal}{Journal of Physics B: Atomic, Molecular and Optical
  Physics} \textbf{\bibinfo{volume}{44}}, \bibinfo{pages}{154013}
  (\bibinfo{year}{2011}).

\bibitem[{\citenamefont{Johansson et~al.}(2012)\citenamefont{Johansson, Nation,
  and Nori}}]{bib:qutip}
\bibinfo{author}{\bibfnamefont{J.}~\bibnamefont{Johansson}},
  \bibinfo{author}{\bibfnamefont{P.}~\bibnamefont{Nation}}, \bibnamefont{and}
  \bibinfo{author}{\bibfnamefont{F.}~\bibnamefont{Nori}},
  \bibinfo{journal}{Computer Physics Communications}
  \textbf{\bibinfo{volume}{183}}, \bibinfo{pages}{1760} (\bibinfo{year}{2012}).

\bibitem[{\citenamefont{Rivas et~al.}(2014)\citenamefont{Rivas, Huelga, and
  Plenio}}]{bib:nmreport}
\bibinfo{author}{\bibfnamefont{A.}~\bibnamefont{Rivas}},
  \bibinfo{author}{\bibfnamefont{S.~F.} \bibnamefont{Huelga}},
  \bibnamefont{and} \bibinfo{author}{\bibfnamefont{M.~B.}
  \bibnamefont{Plenio}}, \bibinfo{journal}{Reports on Progress in Physics}
  \textbf{\bibinfo{volume}{77}}, \bibinfo{pages}{094001}
  (\bibinfo{year}{2014}).

\bibitem[{\citenamefont{Addis et~al.}(2015)\citenamefont{Addis, Ciccarello,
  Cascio, Palma, and Maniscalco}}]{bib:manis}
\bibinfo{author}{\bibfnamefont{C.}~\bibnamefont{Addis}},
  \bibinfo{author}{\bibfnamefont{F.}~\bibnamefont{Ciccarello}},
  \bibinfo{author}{\bibfnamefont{M.}~\bibnamefont{Cascio}},
  \bibinfo{author}{\bibfnamefont{G.~M.} \bibnamefont{Palma}}, \bibnamefont{and}
  \bibinfo{author}{\bibfnamefont{S.}~\bibnamefont{Maniscalco}},
  \bibinfo{journal}{New Journal of Physics} \textbf{\bibinfo{volume}{17}},
  \bibinfo{pages}{123004} (\bibinfo{year}{2015}).

\bibitem[{\citenamefont{Pineda et~al.}(2016)\citenamefont{Pineda, Gorin,
  Davalos, Wisniacki, and Garc{\'\i}a-Mata}}]{bib:pineda}
\bibinfo{author}{\bibfnamefont{C.}~\bibnamefont{Pineda}},
  \bibinfo{author}{\bibfnamefont{T.}~\bibnamefont{Gorin}},
  \bibinfo{author}{\bibfnamefont{D.}~\bibnamefont{Davalos}},
  \bibinfo{author}{\bibfnamefont{D.~A.} \bibnamefont{Wisniacki}},
  \bibnamefont{and}
  \bibinfo{author}{\bibfnamefont{I.}~\bibnamefont{Garc{\'\i}a-Mata}},
  \bibinfo{journal}{Physical Review A} \textbf{\bibinfo{volume}{93}},
  \bibinfo{pages}{022117} (\bibinfo{year}{2016}).

\bibitem[{\citenamefont{Poggi et~al.}(2017)\citenamefont{Poggi, Lombardo, and
  Wisniacki}}]{bib:pablo}
\bibinfo{author}{\bibfnamefont{P.}~\bibnamefont{Poggi}},
  \bibinfo{author}{\bibfnamefont{F.}~\bibnamefont{Lombardo}}, \bibnamefont{and}
  \bibinfo{author}{\bibfnamefont{D.}~\bibnamefont{Wisniacki}},
  \bibinfo{journal}{EPL} \textbf{\bibinfo{volume}{118}}, \bibinfo{pages}{20005}
  (\bibinfo{year}{2017}).

\bibitem[{\citenamefont{Breuer et~al.}(2009)\citenamefont{Breuer, Laine, and
  Piilo}}]{bib:nmbreuer1}
\bibinfo{author}{\bibfnamefont{H.-P.} \bibnamefont{Breuer}},
  \bibinfo{author}{\bibfnamefont{E.-M.} \bibnamefont{Laine}}, \bibnamefont{and}
  \bibinfo{author}{\bibfnamefont{J.}~\bibnamefont{Piilo}},
  \bibinfo{journal}{Physical review letters} \textbf{\bibinfo{volume}{103}},
  \bibinfo{pages}{210401} (\bibinfo{year}{2009}).

\bibitem[{\citenamefont{Breuer et~al.}(2016)\citenamefont{Breuer, Laine, Piilo,
  and Vacchini}}]{bib:nmbreuer2}
\bibinfo{author}{\bibfnamefont{H.-P.} \bibnamefont{Breuer}},
  \bibinfo{author}{\bibfnamefont{E.-M.} \bibnamefont{Laine}},
  \bibinfo{author}{\bibfnamefont{J.}~\bibnamefont{Piilo}}, \bibnamefont{and}
  \bibinfo{author}{\bibfnamefont{B.}~\bibnamefont{Vacchini}},
  \bibinfo{journal}{Reviews of Modern Physics} \textbf{\bibinfo{volume}{88}},
  \bibinfo{pages}{021002} (\bibinfo{year}{2016}).

\bibitem[{\citenamefont{Wi{\ss}mann et~al.}(2012)\citenamefont{Wi{\ss}mann,
  Karlsson, Laine, Piilo, and Breuer}}]{bib:wissmann}
\bibinfo{author}{\bibfnamefont{S.}~\bibnamefont{Wi{\ss}mann}},
  \bibinfo{author}{\bibfnamefont{A.}~\bibnamefont{Karlsson}},
  \bibinfo{author}{\bibfnamefont{E.-M.} \bibnamefont{Laine}},
  \bibinfo{author}{\bibfnamefont{J.}~\bibnamefont{Piilo}}, \bibnamefont{and}
  \bibinfo{author}{\bibfnamefont{H.-P.} \bibnamefont{Breuer}},
  \bibinfo{journal}{Physical Review A} \textbf{\bibinfo{volume}{86}},
  \bibinfo{pages}{062108} (\bibinfo{year}{2012}).

\end{thebibliography}

\end{document}